\documentstyle[12pt]{article}
\newcommand{\be}{\begin{equation}}
\newcommand{\ee}{\end{equation}}
\newcommand{\bea}{\begin{eqnarray}}
\newcommand{\eea}{\end{eqnarray}}
\newcommand{\nn}{\nonumber \\}
\newcommand{\p}[1]{(\ref{#1})}

\newcommand\al{\alpha}
\newcommand\bt{\beta}
\newcommand\de{\delta}
\newcommand\ga{\gamma}
\newcommand\ep{\epsilon}

\begin{document}
\vskip1cm
\centerline{\large\bf Diverse PBGS Patterns and Superbranes}
\vskip1cm
\centerline{\bf E. Ivanov}
\bigskip
\centerline{\it Bogoliubov Laboratory of Theoretical Physics,
JINR,}
\centerline{\it 141 980, Dubna, Moscow Region, Russian Federation}

\vskip1cm

\begin{center}
{\begin{minipage}{4.2truein}
                 \footnotesize
                 \parindent=0pt
This is a brief account of the approach to
superbranes based upon the concept of
Partial Breaking of Global Supersymmetry (PBGS).

\par\end{minipage}}\end{center}
 \vskip 2em \par
\noindent{\bf 1. Introduction.}\hskip 1em
The view of superbranes as theories of partial spontaneous breaking of global
SUSY (PBGS) received a considerable attention (see, e.g.,
\cite{bw} - \cite{ik}). The salient feature of this approach is
that it deals with the Goldstone superfields
living on the worldvolume superspace of
unbroken SUSY and accommodating the superbrane
physical degrees of freedom.

It is well known that the superbranes in the Green-Schwarz (GS) formulation
break half of
the target SUSY (see, e.g., \cite{duff}).
Choosing the static
gauge with respect to the worldvolume diffeomorphisms and killing half of
the target $\theta$-coordinates by $\kappa$-symmetry,
one ends up with the physical multiplet which comprises
transverse target bosonic coordinates and the rest of $\theta$-coordinates
(in the case of D-branes and their generalizations, the physical multiplets
can include additional fields). The remaining $\theta$'s are shifted under
half of the target SUSY, which suggests to interpret them as the relevant
Goldstone fermions. Thus one half of the original SUSY, with respect
to which the physical degrees of freedom form a supermultiplet,
is unbroken, while
the other is spontaneously broken, with the physical fermions as the Goldstone
ones.

One can reverse the argument: adopt PBGS as the guiding principle and
deduce superbranes just from it. In doing so, one can take advantage
of the nonlinear realizations method \cite{1}-\cite{3}
which provides the universal framework for treating spontaneously
broken symmetries. In this method, the Goldstone (super)fields are
identified with
the parameters of the coset space of the full (super)symmetry group
over its unbroken
symmetry subgroup. The invariant actions of the Goldstone superfields
constructed
as integrals over the worldvolume superspace are reduced to the
corresponding GS-type actions
after passing to components and eliminating the auxiliary fields.

The nonlinear realizations method was worked out for
constructing nonlinear sigma models of internal symmetries.
It is less known that the bosonic $p$-branes in the physical (or ``static'')
gauge can also be treated in the nonlinear realizations language
\cite{ivan}. To describe in this way some $p$-brane moving
in $D$-dimensional Minkowski space, one should consider
a nonlinear realization of
the relevant Poincar\'e group ${\cal P}_{D} = ISO(1,D-1)$,
such that the residual
unbroken (vacuum stability)
symmetry group is the product of the Poincar\'e group of
$p+1$-dimensional brane worldvolume and the rotation
group of the transverse brane coordinates. The relevant coset manifold is
${\cal P}_D/SO(1,p)\otimes SO(D-p-1)$.
One splits the full $D$ space translation generator
$P_M$, $\;M=0,1, \ldots D-1$, as
\be
P_M \quad \Rightarrow \quad (P_m, \; P_\mu)\;, \;\; m = 0,1,
\ldots p; \;\;\mu = p+1, \ldots D~,
\ee
and associates with $P_m, P_\mu$ the worldvolume coordinate $x^m$
and the Goldstone field $X^\mu(x)$ as the coset parameters
($X^\mu$ becomes the transverse
$p$-brane coordinate). Also, one introduces the Goldstone fields
$\Lambda^{\mu}_{m} (x)$ parametrizing the spontaneously broken
part of the Lorentz
group $SO(1,D-1)$ (with the generators $L_{\mu}^m$):
\be
P_m \;\Rightarrow \; x^m~, \quad  P_\mu \;\;\Rightarrow \;\;
X^\mu (x)~, \quad
L_{\mu}^m \;\Rightarrow \; \Lambda^{\mu}_m (x)~.
\ee
The application of the Cartan forms techniques augmented with
some extra covariant constraints (the inverse Higgs \cite{hig}) gives rise
to the following minimal invariant action
\be
S_{br} \sim \int d^{p+1} x \left(\sqrt{(-1)^p\,g} -1 \right), \;
g = \mbox{det}\,(\eta_{mn}
-\partial_mX^\mu \partial_n X^\mu)~.
\label{ng}
\ee
It is the static gauge form of the $p$-brane Nambu-Goto (NG) action.

As an illustration, let us consider the simplest example of
the massive particle ($0$-brane) in
$D=2$. The $D=2$ Poincar\'e group ${\cal P}_{(2)}$ involves
two translation generators $P_0, P_1$ and the
$SO(1,1)$ Lorentz generator $L$, with the only non-vanishing commutators
\be
[\,L, P_0\,] = iP_1~, \quad  [\,L, P_1\,] = i P_0~.
\ee
In accord with the said above, we should construct
a nonlinear realization of ${\cal P}_{(2)}$, with the
one-dimensional ``Poincar\'e group'' generated
by the generator $P_0$ as the stability subgroup.
Thus we are led to place all generators into the
coset (in this particular case it coincides with the full group):
\be
G = \mbox{e}^{itP_0} \, \mbox{e}^{iX(t)P_1}\, \mbox{e}^{i\Lambda(t)L}~.
\label{toycos}
\ee
Here the wordline evolution parameter $t$ (time) is the coset
coordinate associated
with $P_0$, and the Goldstone fields are associated with the rest
of generators.
The group ${\cal P}_{(2)}$ acts as left shifts of $G$. The Cartan forms
\be
G^{-1}\,d G = i\,\omega_t\, P_0 + i\,\omega_1\,P_1 + i\omega_L\,L~,
\label{toyf}
\ee
\bea
&& \omega_t = \sqrt{1 + \Sigma^2}\,dt + \Sigma \,dX~, \quad
\omega_1 = \sqrt{1 + \Sigma^2}\,dX + \Sigma \,dt~, \nn
&& \omega_L= {1\over \sqrt{1 + \Sigma^2}}\,d\Sigma~, \qquad
\Sigma \equiv \mbox{sh}\,\Lambda
\label{om}
\eea
by construction are invariant under this left group action.
Next, we observe that the Lorentz Goldstone field $\Sigma (t)$
can be traded for
$\dot{X}(t)$ by the inverse Higgs \cite{hig} constraint
\be
\omega_1 = 0 \quad \Rightarrow \quad \Sigma =
- {\dot{X}\over \sqrt{1-\dot{X}^2}}~.
\label{ih}
\ee
This constraint is covariant since $\omega_1$ is the group invariant (in the
generic case, the coset Cartan forms undergo homogeneous rotation
in their stability subgroup indices). Thus the obtained expression
for $\Sigma$ possesses correct
transformation properties. Substituting it into the remaining
Cartan forms we find
\be
\omega_0 = \sqrt{1 - \dot{X}^2}\,dt~, \quad \omega_L =
\sqrt{1 - \dot{X}^2}\,
\frac{d}{dt}\left(\frac{\dot{X}}{\sqrt{1 - \dot{X}^2}}\right)dt~.
\ee
The simplest invariant action, the covariant length
\be
S = \int \omega_0 = \int dt \sqrt{1 - \dot{X}^2}~,
\label{toyact}
\ee
is recognized, up to a renormalization factor of the dimension of mass,
as the action of $D=2$ massive particle in the static gauge $X^0 (t) = t$.
The equation of motion for $X(t)$ can also be given the manifestly
covariant form
\be
\omega_L = 0~.
\ee
Actually, we could start from the action \p{toyact} with the
original expression \p{om}
for $\omega_0$ and reproduce \p{ih} as the algebraic equation of
motion for $\Sigma(t)$.
It is important to realize that the correct form of the action
is recovered just because we have included the Lorentz Goldstone field
into the coset.

In the generic case of $p$-brane in $D$-dimensional Minkowski space
the construction is analogous. One equates to zero
the Cartan forms $\omega^\mu$ associated with the transverse
translation generators, which
is once again the manifestly covariant constraint, and in this way
eliminates the Lorentz Goldstone
fields $\Lambda^{\mu m} (x)$:
\be
\Lambda^{\mu}_m \sim \partial_m X^\mu~.
\ee
Substituting these expressions into the Cartan forms $\omega^m$ which
are covariant
differentials of $x^m$, one finds that the invariant volume of $x$-space,
i.e. the integral of the
external product of these 1-forms, is just the $p$-brane static
gauge NG action \p{ng}.

\vspace{0.2cm}

\noindent{\bf 2. Superbranes from PBGS.} The PBGS approach is
the generalization
of the above nonlinear realization view of branes to the superbranes.
Though originally \cite{bw} the phenomenon of partial
breaking of global SUSY
($N=2$ down to $N=1$ in $D=4$) was studied without any reference
to superbranes, the
subsequent study \cite{hp}-\cite{ik} revealed
the profound relationship between both
concepts. It was shown in \cite{hp} that $N=1, D=4$ superstring
in the static gauge
can be understood as the theory of partial breaking
of $N=1\,, D=4$ SUSY to its
$N=(2,0)\,, d=2$ subgroup. The very existence of self-consistent
GS type actions for
superbranes, with the appropriate fermionic $\kappa$-symmetry,
was inferred in the
pioneering paper \cite{hlp} from the study of the partial breaking
$N=(1,0)\,, D=6 \;\Rightarrow \;N=1\,, d=4$.
The same PBGS pattern was treated in
\cite{bg1} from the $D=4$ perspective as the partial breaking
of $N=2$ SUSY with two central
charges down to $N=1$ one, with the systematic use of the nonlinear
realizations and inverse Higgs effect techniques. It was shown that
the self-consistent theory can be constructed in terms of the (covariantly)
chiral and antichiral bosonic $N=1$ superfields which are the Goldstone ones
corresponding to the central charges (that is, to the translation
operators in fifth and sixth
directions from the $D=6$ viewpoint). The fermionic Goldstone
superfields associated
with the spontaneously broken supertranslation generators
are covariantly expressed in terms of these
basic Goldstone superfields by the covariant constraints of the type \p{ih}.
The relevant Goldstone superfields action, a nonlinear generalization of the
standard free action of $N=1$ chiral superfields, was constructed in
\cite{{bg3},{rt}}. In components,
after eliminating the auxiliary fields, it yields just
the static gauge form of the GS action
for the $N=(1,0)\,, D=6$ 3-brane in a flat background.
An interesting new phenomenon was discovered
in \cite{bg2}. It turned out that the same SUSY admits several
different PBGS options depending
on into which multiplet of unbroken SUSY one embeds the Goldstone fermion.
In the $N=2\,\;\Rightarrow \;N=1\,, D=4$ case, instead of placing
this field into
the chiral $N=1$ multiplet, one can place it into
the abelian vector $N=1$ multiplet as a ``photino'',
by imposing the appropriate covariant
constraints on the Goldstone fermionic superfield which
generalize the Bianchi identities for
the flat $N=1$ Maxwell superfield strength.
The relevant invariant action, on the one hand,
is $N=2$ extension of the Born-Infeld (BI) action with the hidden,
nonlinearly realized half of supersymmetry,
and, on the other, is (in components) a gauge-fixed form of
the GS action of the ``space-time filling''
D3-brane. One more option is to embed the Goldstone fermion into
the $N=1$ tensor (or linear)
multiplet \cite{{bg3},{rt},{gpr}}. The emerging brane
is the super ``L3-brane'' in $D=5$
(in terminology of \cite{Lbr}), it is related to the
$N=(1,0)\,, D=6\;$ 3-brane via
the familiar duality between $N=1$ tensor and chiral superfields.

The study of the PBGS patterns corresponding to partial breaking
of SUSY with
16 supercharges ($N=1\,, D=10$; $N=2, D=6$; $N=4\,, D=4$; ...)
was initiated in our
works \cite{bik1,bik2}. Let us dwell on this and related subjects
in some detail.

\vspace{0.2cm}
\noindent{\bf 3. Hypermultiplet as a Goldstone superfield.}
To describe the $1/2$ breaking of Poincar\'e SUSY with 16 supercharges, it
is natural to start from the maximally symmetric situation where the broken
and unbroken SUSY ``live'' as simple ones. This amounts to
considering the PBGS option $N=1,\,D=10 \;\Rightarrow \; N=(1,0), \,d=6$
(we could equally choose $N=(0,1)$). From the $d=6$ perspective,
$N=1, \,D=10$ is a central extension of $N=(1,1)$:
\be
N=1,\; D=10\;\;\; \quad \propto \quad
\left\{ Q^i_\alpha, P_{\alpha\beta}, S^{\beta a}, Z^{ia}
\right\}~, \label{setsus}
\ee
where $\alpha, \beta = 1,...,4~; \; i = 1,2~; \; a = 1,2
$
are, respectively, the $Spin(1,5)$ indices and the
$SU(2)$ doublet indices. The basic anticommutation relations read
\be
\left\{ Q_{\al}^i,Q_{\bt}^j\right\}=\ep^{ij}P_{\al\bt}\; ,\;
\left\{ Q_{\al}^i,S^{a\bt}\right\} = \de_{\al}^{\bt}Z^{ia}\; , \;
\left\{ S^{a\al},S^{b\bt}\right\} = \ep^{ab}P^{\al\bt} \;. \label{susy}
\ee
One also must add generators of the $D=10$ Lorentz
group $SO(1,9)\propto \left\{ M_{\al\bt\;\ga\de},\;
T^{ij}, \; T^{ab}, \; K_{ia}^{\al\bt}\right\}$, with $M$ and $T$
generating the $d=6$ Lorentz
group $SO(1,5)$ and $R$-symmetry $SU(2)\times SU(2)$.

We wish $N=(1,0), d=6$ SUSY $\propto \left\{Q^i_\alpha,
P_{\alpha\beta}, T^{ij},
T^{ab},
M_{\al\bt\;\ga\de}
\right\}$ to be unbroken. Like
in the bosonic case of Sect.1, we are led to place the generators
$Q^i_\alpha,
P_{\alpha\beta}, S^{\alpha a}, Z^{ia},
K^{ia}_{\alpha\beta}$ into the coset and to treat the coset parameters
associated with two first generators as the coordinates of
$N=(1,0), \,d=6$ superspace, while the remaining ones as
the Goldstone superfields:
\bea
&&Q^i_\alpha \Rightarrow \theta^\alpha_i~, \quad P_{\alpha\beta}
\Rightarrow x^{\alpha\beta}~. \nn
&&S^{\alpha a} \Rightarrow \Psi_{\alpha a}(x, \theta)~, \;
Z^{ia} \Rightarrow q_{ia}(x,\theta)~, \; K^{ia}_{\alpha\beta}
\Rightarrow \Lambda^{\alpha\beta}_{ia}(x,\theta)~. \label{Golddef}
\eea
The coset element $G$ in the exponential parametrization is constructed like
in eq. \p{toycos}.
The Cartan forms are defined by
\be
G^{-1}d G = \Omega_Q + \Omega_P + \Omega_Z + \Omega_S + \Omega_K + \ldots ~,
\ee
where the subscripts denote the relevant generators. The forms $\Omega_{Q,P}$
are the covariant differentials of the superspace coordinates,
$\Omega_{Z,S,K}$ are those of Goldstone superfields,
they are homogeneously transformed under the supergroup left shifts.
We shall be interested in the linearized structure of
$\Omega_Z = \Omega_Z^{ia}\;Z_{ia}$ \cite{bik1,bik2}:
\be
\Omega_Z^{ia} = dq^{ia} + 2\Lambda^{ia}_{\al\bt}\,dx^{\al\bt} +
\psi^a_\alpha d\theta^{i\al} + \ldots ~.
\ee
Comparing it with the form $\omega_1$, eq. \p{om} of
the toy example of Sect. 1,
we observe that both forms have a similar structure. Hence, the
superfields $\Lambda, \Psi $ can be expressed through the
basic Goldstone superfield $q^{ia}$ from the inverse
Higgs constraint analogous to \p{ih}
\bea
\Omega_Z = 0 \;\Rightarrow\; \Lambda^{ia}_{\al\bt} &=&
\nabla_{\al\bt}q^{ia} = \partial_{\al\bt}q^{ia}
+ \ldots~, \label{basconstr} \\
\Psi^a_\al &=& {1\over 2}\nabla^k_\al q^a_k =
{1\over 2}D^k_\al q^a_k + \ldots ~. \nonumber
\eea
Here $D^i_\al = \partial/\partial \theta^\al_i -
{1\over 2}\theta^{i\bt}\partial_{\bt\al}$ and dots stand for nonlinear
terms. Thus, $q^{ia}$ is the essential Goldstone
superfield, analogue of $X(t)$ in the $D=2$ example. Its first bosonic
component $q^{ia}(x)$  parametrizes the transverse directions in the
$D=10$ Minkowski space, while the physical fermionic component is
the Goldstino
related to the spontaneously broken supertranslations (with the generator
$S^{a\al}$). This field content is that of the scalar $N=1\,, D=10$ 5-brane.

The constraint \p{basconstr} differs in a few important aspects
from \p{ih} and the similar inverse Higgs constraint
considered in ref. \cite{bg1}. It not only eliminates the redundant
Goldstone superfields, but also implies the differential
constraint for $q^{ia}(x,\theta)$
\be
\nabla^{(k}_\al q^{i)a} = D^{(k}_\al q^{i)a} + \ldots = 0~. \label{dyn}
\ee
It is just the nonlinear, ``brane'' generalization of the standard
hypermultiplet constraint \cite{fs}. The latter reduces the
field content of $q^{ia}$ to the set of (8+8) components and puts
them on shell:
\bea
&&q^{ia}(x, \theta) \;\;\Rightarrow \;\; \phi^{ia}(x) + \theta^{\alpha i}
\psi_\alpha^a (x) + x\mbox{-derivatives}~, \label{reduct} \\
&& \Box \phi^{ia}(x) = 0~, \quad \partial^{\alpha\beta}\psi^a_\bt = 0~.
\eea

Thus eq. \p{dyn} describes the on-shell dynamics of
$N=1, D=10$ 5-brane as the natural generalization of the free hypermultiplet
dynamics. Since the off-shell superfield action for the latter can be
constructed only in harmonic superspace \cite{gikos}, it is reasonable
to expect that
there exists a brane generalization of the hypermultiplet harmonic superspace
action. It would be $N=(1,0), d=6$ (or $N=2, d=4$ after dimensional reduction)
analogue of the $N=1$ Goldstone superfield action of refs. \cite{bg3,rt}.
No systematic
recipes are known so far
for constructing PBGS actions. At present we are aware only of the
appropriate fourth-order correction to the free hypermultiplet action.
Even for the dimensionally-reduced cases, including the simplest case of
$N=2\,, D=5$ superparticle, the full action is rather difficult to find.
For the superparticle case it is known up to
the sixth order in fields \cite{bik2}:
\be
S^q_{br} = \int d\zeta^{(-4)} q^{+}_aD^{++}q^{+a} +
 \al\int d Z (A^2  + 2\al\,A B^{++}B^{--} + \ldots) \label{d14}
\ee
where
$$
A = q^{+}_aD^{--}q^{+a}~, \quad B^{++} = q^+_a\partial_tq^{+a}~,
$$
$q^{+a}, D^{\pm\pm}$ are the $d=1$ reduction of the analytic
hypermultiplet superfield
and harmonic derivatives \cite{gikos}, $B^{--}$ is defined by
$$
D^{++}B^{--} - D^{--}B^{++} = 0
$$
and $\alpha$ is a coupling constant. It would be tempting
to find out the geometric principle allowing
to restore the whole action. Let us now apply to a simpler PBGS case
where the complete off-shell action can be constructed \cite{ik}.

\vspace{0.2cm}

\noindent{\bf 4. N=1, D=4 supermembrane.} This case corresponds
to the partial breaking
of $N=1, D=4$ SUSY to $N=1, d=3$ one. The $N=1, D=4$ superalgebra
in the $d=3$ notation
reads
\be
\{ Q_{a},Q_{b}\}=P_{ab}, \{ Q_{a},S_{b}\} = \ep_{ab}Z,
\{ S_{a},S_{b}\} = P_{ab},\; a,b = 1,2~. \label{susy34}
\ee

The $d=3$ momentum operator $P_{ab}$ together with the central
charge $Z$ form
the $D=4$ translation operator. The vacuum stability subalgebra consists
of the $N=1, d=3$ superalgebra (generators $Q_a$ and $P_{ab}$) and that of
the $d=3$ Lorentz group $SO(1,2)$. The second SUSY, the central charge $Z$ and
the generators $K_{ab}$ of the coset $SO(1,3)/SO(1,2)$ define
spontaneously broken
symmetries. The appropriate coset parameters are introduced as
\be
Q_a \Rightarrow \theta^a, P_{ab}
\Rightarrow x^{ab},  S_a \Rightarrow \psi^a(x,\theta),
Z \Rightarrow q(x,\theta),  K_{ab} \Rightarrow \lambda^{ab}(x,\theta).
\label{Golddef1}
\ee

After construction of the relevant Cartan forms and covariant elimination of
the Goldstone superfields $\psi^a(x,\theta),\lambda^{ab}(x,\theta)$
by the appropriate inverse Higgs
constraints, one is left with $q(x,\theta)$ as the only unremovable Goldstone
superfield. Its physical fields (one boson and two fermions)
parametrize one transverse bosonic and two fermionic directions in
$N=1, D=4$ superspace,
the auxiliary bosonic field also
admits a nice geometric interpretation of the Goldstone field for the
spontaneously broken $U(1)$ automorphism group of
$N=1, D=4$ superalgebra (``$\gamma^5$
invariance'').
The physical field content of $q$ coincides with that of $N=1, D=4$
supermembrane.

The inverse Higgs conditions in this case do not imply the equation
of motion for $q$.
However, the dynamical equation for $q$ turns out to admit,
like in the $D=2$ particle
example, a covariant representation as the vanishing of
the covariant $d\theta$
projection of the coset Cartan form $\omega_S^a$ associated
with the spontaneously broken
SUSY generator $S_a$ \cite{ik}. We have no place to present details.
Let us explain
how to construct the off-shell action for this case.

Like in other PBGS cases, the nonlinear realizations approach
on its own provides no clear
recipe how to construct such an action. This becomes possible using
the trick similar to the one exploited in \cite{bg2}. Namely,
let us start from a {\it linear} realization of $N=1, D=4$ SUSY
in terms of $N=1, d=3$ superfields $\Phi, \xi_a \equiv D_a\rho $
with the following transformation
rules under the second SUSY:
\be
\delta \rho = \theta^a\eta_a - 2 D^a \Phi\eta_a~, \; \delta \Phi =
{1\over 2} \eta^aD_a\rho~,
\label{transmem}
\ee
where $\eta_a$ is the transformation parameter and $D_a$ is the flat
$N=1, d=3$ spinor covariant
derivative, $\{D_a, D_b\} = \partial_{ab}$.
It is easy to check that the closure of these transformations
and those of manifest $N=1, d=3$ SUSY
is just the superalgebra \p{susy34}, with $Z$ realized as
a pure shift of $\rho$.
The transformation of $\xi_a = D_a\rho$ starts with $\eta_a$,
suggesting
the interpretation of this superfield as the linear realization
Goldstone fermion.
After some work one finds that $\Phi$ can be covariantly
expressed in terms of $\xi_a$ as follows
$$
\Phi = {1\over 2} \frac{\xi^2}{1 + \sqrt{1 + D^2\xi^2}}~.
$$
Recalling the transformation law of $\Phi$, one finds that the integral
\be\label{action1}
S= \int d^3 x d^2 \theta \Phi \equiv \frac{1}{2}
\int d^3 x d^2 \theta
 \frac{ \xi^2}{1+\sqrt{1+D^2\xi^2}}, \quad
\xi^a = D^a\,\rho~,
\ee

is invariant under the hidden SUSY transformations as well
as $D=4$ Poincar\'e
translations and so it can be identified with
the sought $d=3$ worldvolume superspace action of $N=1, D=4$ supermembrane.
Indeed, it is easy to find that the bosonic core of this action is just
the static gauge membrane NG action:
\be  \label{NG1} S= \int d^3x \left( 1 -
\sqrt{1-\frac{1}{2}\partial q \cdot \partial q} \right)\; .
\ee
One can find the equivalence field redfinition relating $\xi_a$ to the
nonlinear realization Goldstone fermion $\psi_a$ and $\rho$ to $q$.
Also, the equations
of motion following from the action \p{action1} can be shown
to be equivalent to
those conjectured at the level of the Cartan forms. This implies the presence
of hidden $SO(1,4)$ Lorentz symmetry in the action.  It still remains to prove
the precise equivalence of this action to the GS one (e.g., along the lines of
ref. \cite{town}) and the action proposed in the superembedding approach
\cite{Lbr}.
Note an interesting peculiarity: one can add
to the lagrangian in
\p{action1} the ``cosmological''  term $\sim \rho $.
This term is invariant under the second SUSY (up to surface terms) and the
shifts $\rho \rightarrow \rho + const$. Adding it changes the equation
of motion for the auxiliary field and so can influence the structure
of the component action (without this term, the auxiliary field is vanishing
on shell). The pure physical boson part of the action is always
given by \p{NG1}.

Besides a
scalar multiplet $\rho$ (or $q$),  we can choose a vector $N=1, d=3$
multiplet as the Goldstone one (like in \cite{bg2}). It is represented
by $N=1$ spinor superfield strength $\mu_a$ obeying the constraint:
\be\label{cc1} D^a\mu_a=0~.
\ee
It leaves in $\mu_a$ the first fermionic
component together with the
divergenceless vector $F_{ab}\equiv D_a\mu_b|_{\theta=0}$
(just the gauge field strength). Due to the vector-scalar $d=3$ duality,
the superfield $\mu_a$ is expected to describe
a D2-brane which is dual to the supermembrane.

The relevant action can be found using the previous trick. One can
extend $\mu^a$ to
the $N=1, D=4$ multiplet $(\mu^a, \phi)$ with the following
transformation rules under
the second SUSY
\be \label{tr2} \delta \mu_a = \eta_a - D^2 \phi
\eta_a + \partial_{ab} \phi \eta^b \; , \qquad \delta \phi =
\frac{1}{2} \eta^a \mu_a \;.
\ee
Thus $\mu^a$ can also be interpreted as the linear realization Goldstone
fermionic superfield. The following expression for $\phi$
\be
\phi= \frac{1}{2}\; \frac{\mu^2 }{1+\sqrt{1-D^2\mu^2}}
\ee
can be checked to be consistent with \p{tr2}. Then the action
\be\label{actionm}
S=-\int d^3x d^2\theta\, \phi = - \frac{1}{2}\,\int d^3x d^2\theta
\,\frac{\mu^2 }{1+\sqrt{1-D^2\mu^2}}
\ee
is invariant under the second SUSY in virtue of the transformation rule of
$\phi$ \p{tr2} and the constraint \p{cc1}. This nonlinear generalization of
the standard $N=1, d=3$ abelian vector multiplet action $\sim \mu^2$ involves
as its bosonic core the $d=3$ BI action
\bea
&& S= \int d^3x \left( \sqrt{1+2F^2}-1 \right) \;, \label{bm} \\
&& \partial^{ab}F_{ab}=0 \quad \rightarrow \quad
     F_{ab}=\partial_{ac}G^c_b+\partial_{bc}G^c_a \; .
\eea
So it is $N=2$ extension of the $d=3$ BI action with nonlinearly
realized second SUSY. It can be regarded as the worldvolume
superfield action of the space-time filling D2-superbrane in a flat
background. It is easy to prove its dual equivalence to the action \p{action1}
by inserting the constraint \p{cc1} in it with the
Lagrange multiplier $\rho$ and integrating out $\mu^a$ \cite{ik}.

Finally, let us show how the duality between the bosonic NG and BI
actions \p{NG1}, \p{bm} can be recovered in the nonlinear realizations
approach
of Sect.1. It is easy to
find that in the case of nonlinear realization of ${\cal P}_{(4)}$
in the coset ${\cal P}_{(4)}/SO(1,2)$, which corresponds just to membrane
in $D=4$,
the covariant differentials of $x^m$ read
\be
\omega^m =
dx^m + 2 \frac{\lambda^m}{1-2\lambda^2}\left(2\lambda_n
+ \partial_n q\right) dx^n \equiv \omega^m_n dx^n~,
\label{ommemb}
\ee
where $\lambda^m(x)$ is the Lorentz $SO(1,3)/SO(1,2)$ Goldstone field in
the appropriate parametrization and $q(x)$ is the transverse
coordinate of membrane. We could eliminate $\lambda^m$ by the inverse Higgs
constraint but in the present case it is instructive to reproduce
it as the equation of motion for $\lambda^m$. The minimal
invariant action constructed as the covariant worldvolume
\be
S_{mem} = \int d^3x\,\mbox{det}\,\omega^m_n~,
\ee
up to a normalization factor and constant shift, is
\be
S_{mem} = \int d^3x \frac{1}{1 - 2\lambda^2}\left(2\lambda^2 +
\lambda\,\partial q \right)~.
\label{omS}
\ee
Varying $\lambda^m$ yields the inverse Higgs expression for it
\be
\lambda_m = -{1\over 2}\,{\partial_m q\over
1 +\sqrt{1-{1\over 2} (\partial q)^2}}~.
\ee
After this \p{omS} takes the standard NG form \p{NG1}.
On the other hand, one can treat $q$ in
\p{omS} as the Lagrange multiplier for the differential constraint on
$\lambda^m$:
\be
\partial_m F^m = 0~, \quad F^m \equiv 2\,{\lambda^m\over 1 - 2\lambda^2}~.
\label{bianchi}
\ee
Expressing $\lambda^2$ through $F^m$
$$
\lambda^2 = {1\over 4F^2}\left(1-\sqrt{1 +  2F^2} \right)^2,
$$
one reduces \p{omS} just to the BI
form \p{bm}
\be
S_{mem} \sim \int d^3 x \left(\sqrt{1 + 2 F^2} - 1 \right)~.
\ee
This simple consideration shows that within the nonlinear realization
approach the $d=3$ Maxwell field strength entering the $d=3$ BI action
acquires the nice geometric interpretation as the Goldstone field
representing the Lorentz coset $SO(1,3)/SO(1,2)$, while the BI action itself
can be algorithmically derived as the action dual to the static gauge
membrane NG action.

\vspace{0.2cm}
\noindent{\bf 5. Concluding remarks.} The geometric PBGS approach
can be thought of as a useful and viable alternative to the
standard GS description of superbranes. Its main merit is
that it gives {\it manifestly} worldsurface supersymmetric
off-shell superfield actions. Leaving aside
such important conceptual questions as whether it can be helpful
for quantization, etc, we list here a few more modest problems
solving which could extend its range of applicability.

First of all, it is desirable to develop convenient general recipes of
constructing superfield PBGS actions similar to those provided by
the nonlinear realizations for the internal symmetries
sigma models. At present, constructing such actions is an art to
some extent.

It is important to learn how to construct self-consistent
PBGS actions on non-trivial
bacgrounds involving the worldvolume and target supergravity and
super Yang-Mills fields.

At last, it seems interesting to set up PBGS actions
for the systems with $1/4$ and other exotic partial breaking options.
Some progress in this direction is reported in the contribution by
Delduc, Krivonos and myself in this Volume.

\vspace{0.2cm}

\noindent{\bf Acknowledgements.}\hskip 1em
I thank Jerzy Lukierski
for inviting me to give this contribution.
This work was supported in part by grants
 RFBR-CNRS 98-02-22034,
RFBR 99-02-18417, INTAS-96-0538, INTAS-96-0308
and Nato Grant No.PST.CLG 974874.


\begin{thebibliography}{99}
\bibitem{bw} J. Bagger, J. Wess, Phys. Lett. {\bf B 138} (1984) 105.
\bibitem{hp} J. Hughes, J. Polchinski, Nucl. Phys. {\bf B 278} (1986) 147.
\bibitem{hlp} J. Hughes, J. Liu, J. Polchinski, Phys. Lett. {\bf B 180}
(1986) 370.
\bibitem{town} A. Achucarro, J. Gauntlett,  K. Itoh, P.K. Townsend,
Nucl. Phys. {\bf B 314} (1989) 129.
\bibitem{bg1} J. Bagger, A. Galperin, Phys. Lett. {\bf B 336}
(1994) 25.
\bibitem{bg2} J. Bagger, A. Galperin, Phys. Rev. {\bf D 55} (1997)
1091.
\bibitem{bg3} J. Bagger, A. Galperin, Phys. Lett. {\bf B 412} (1997) 296.
\bibitem{rt} M. Ro\v{c}ek, A. Tseytlin, Phys. Rev. {\bf D 59} (1999) 106001.
\bibitem{gpr}
F. Gonzalez-Rey, I.Y. Park, M. Ro\v{c}ek, Nucl. Phys. {\bf B 544}
(1999) 243.
\bibitem{bik1}
S. Bellucci, E. Ivanov, S. Krivonos,
Fortschr. f. Phys. {\bf 48} (2000) 19.
\bibitem{bik2} S. Bellucci, E. Ivanov, S. Krivonos, Phys. Lett. {\bf B 460}
(1999) 348.
\bibitem{ik} E. Ivanov, S. Krivonos, Phys. Lett. {\bf B 453}
(1999) 237.
\bibitem{duff} M.J. Duff, ``Supermembranes'', {\tt hep-th/9611203}.
\bibitem{1} S. Coleman, J. Wess, B. Zumino, Phys. Rev. {\bf 177} (1969)
2239;\\
C. Callan, S. Coleman, J. Wess, B. Zumino, Phys. Rev. {\bf 177} (1969)
2247.
\bibitem{2} D.V. Volkov, Sov. J. Part. Nucl. {\bf 4} (1973) 3.
\bibitem{3} V.I. Ogievetsky, Proceedings of X-th Winter School of
Theoretical
Physics in Karpacz, Vol.1. p. 227 (Wroclaw, 1974).
\bibitem{ivan} E. Ivanov, unpublished.
\bibitem{hig} E.A. Ivanov, V.I. Ogievetsky, Teor. Mat. Fiz. {\bf 25}
(1975) 164.
\bibitem{Lbr}
P.S. Howe, O. Raetzel, E. Sezgin, JHEP {\bf 9808} (1998) 011;\\
P.S. Howe, O. Raetzel, I. Rudychev, E. Sezgin,
Class. Quant. Grav. {\bf 16} (1999) 705.
\bibitem{fs}
M.F. Sohnius, Nucl. Phys. {\bf B 138} (1978) 109.
\bibitem{gikos} A. Galperin, E. Ivanov, S. Kalitzin, V. Ogievetsky,
E. Sokatchev, Class. Quant. Grav. {\bf 1} (1984) 469.

\end{thebibliography}
\end{document}